\documentclass[twocolumn,aps,prl,amsmath,amssymb,color,longbibliography,superscriptaddress]{revtex4-2}
\usepackage{stmaryrd}
\usepackage{amsmath}
\usepackage{amssymb}
\usepackage{graphicx}
\usepackage{dcolumn}
\usepackage{bm}
\usepackage{tabularx}
\usepackage{diagbox}
\usepackage{adjustbox}
\usepackage{color}
\usepackage{xcolor}
\usepackage{colortbl}
\usepackage{booktabs}
\usepackage{multirow}
\usepackage{enumitem}
\usepackage[normalem]{ulem}
\usepackage[colorlinks=true,linkcolor=blue,citecolor=blue,urlcolor=blue]{hyperref}

\usepackage[none]{hyphenat} 
\begin{document}
\begin{titlepage}
\title{Layer-Resolved Nonlinear Optics in Finite-Thickness Two-Dimensional Systems}
\author{Liangting Ye}
\affiliation{Beijing Computational Science Research Center, Beijing 100193, China}
\author{Chengzhi Wu}
\affiliation{Beijing Computational Science Research Center, Beijing 100193, China}
\author{Zeyu Jiang}
\email{jiangzy@csrc.ac.cn}
\affiliation{Beijing Computational Science Research Center, Beijing 100193, China}
\author{Bing Huang}
\email{Bing.Huang@csrc.ac.cn}
\affiliation{Beijing Computational Science Research Center, Beijing 100193, China}
\affiliation{School of Physics and Astronomy, Beijing Normal University, Beijing 100875, China}

\begin{abstract}
Nonlinear optical (NLO) responses in two-dimensional quantum-confined systems are typically described within bulk-based frameworks as macroscopic spatial averages. In finite-thickness van der Waals multilayers directly relevant to nanoscale devices, this picture substantially breaks down. Here, we establish a general symmetry-based framework for classifying second-order NLO responses in multilayers. We reveal a layer-resolved organization into skin, weak-skin, and hidden effects governed by local symmetry and stacking order. First-principles calculations for both nonmagnetic and spin-polarized systems confirm our predictions, demonstrating that stacking alone suffices to dramatically reshape both the spatial pattern and magnitude of the NLO response, a phenomenon not explainable within standard bulk theory. Our results establish stacking geometry as an effective knob for engineering surface-selective NLO responses in layered materials.
\end{abstract}

\maketitle
\draft
\vspace{2mm}
\end{titlepage}

\textbf{\emph{Introduction.}} 
Second-order nonlinear optical (NLO) responses, such as second-harmonic generation (SHG) and the bulk photovoltaic effect (BPVE), are governed by spatial symmetry of crystal, making them powerful probes of light-matter interactions in solids\cite{Boyd2023, PhysRevLett.7.118, fridkinBulkPhotovoltaicEffect2001b, liMoS2hBN2013, tan2016enhancement, sun2019giant, dirnbergerMagnetoopticsVanWaals2023a, jamesRecentAdvancementsOptical2021b, ma2019observation, PhysRevLett.93.083904}. Two-dimensional (2D) van der Waals (vdW) materials, owing to their high tunability and flexible multilayer assembly, provide a promising platform for exploring and engineering NLO responses.\cite{sunOpticalModulators2D2016, zotevNanophotonicsMultilayerVan2025}. Their weak interlayer bonding enables diverse stacking registries\cite{PhysRevLett.130.146801, shao2023engineering, hsuSecondHarmonicGeneration2014a, shanStackingSymmetryGoverned2018, foxStackingOrderEngi2024}, giving rise to local symmetry environments that differ from both infinite bulk and isolated monolayers---a feature inaccessible to traditional bulk-based descriptions.

In finite-thickness vdW multilayers directly relevant to nanoscale devices, NLO responses inherently go beyond the widely used macroscopic description based on spatial averaging\cite{zhouSkinEffectNonlinear2024, Mu2023MagneticProximityBPVE, Zhu2023SnP2Se6SHG,Majerus2023OpticalModeling2D}. Instead, they can exhibit strongly nonuniform and coherent spatial organization. A recent example has revealed layer-resolved distributions of NLO responses in layered antiferromagnets with inversion-time-reversal symmetry\cite{zhouSkinEffectNonlinear2024}, in sharp contrast to the macroscopic bulk picture. This finding highlights the need for a universal understanding of how NLO responses are organized in finite-thickness systems. In particular, a unified framework that captures their layer-resolved spatial structure and its relation to stacking-dependent symmetry is still lacking. This motivates us to establish a complete theory connecting stacking-controlled local symmetry with layer-resolved NLO responses in finite-thickness multilayers.

In this Letter, we develop a symmetry-based theory built on the inversion ($\mathcal{P}$) and parity-time $\mathcal{PT}$ symmetries to classify layer-resolved second-order NLO responses in 2D nonmagnetic and spin-polarized multilayers. Taking the point-group symmetry of individual layers as input, the layer-resolved NLO distributions are directly determined by the stacking order. Incorporating the finite range of interlayer couplings, the framework further distinguishes between regions with vanishing and nonvanishing NLO responses across the multilayer. Within this classification, the spatial profiles of NLO coefficients fall into three distinct categories---skin, weak-skin, and hidden effects---together with a trivial featureless distribution. Guided by this framework, we perform first-principles calculations on two representative systems, blue phosphorus and A-type antiferromagnetic CrI$_3$, demonstrating that stacking alone is sufficient to largely modulates the layer-resolved NLO responses.

\begin{figure*}[!t]
	\centering
	\includegraphics[width=\textwidth]{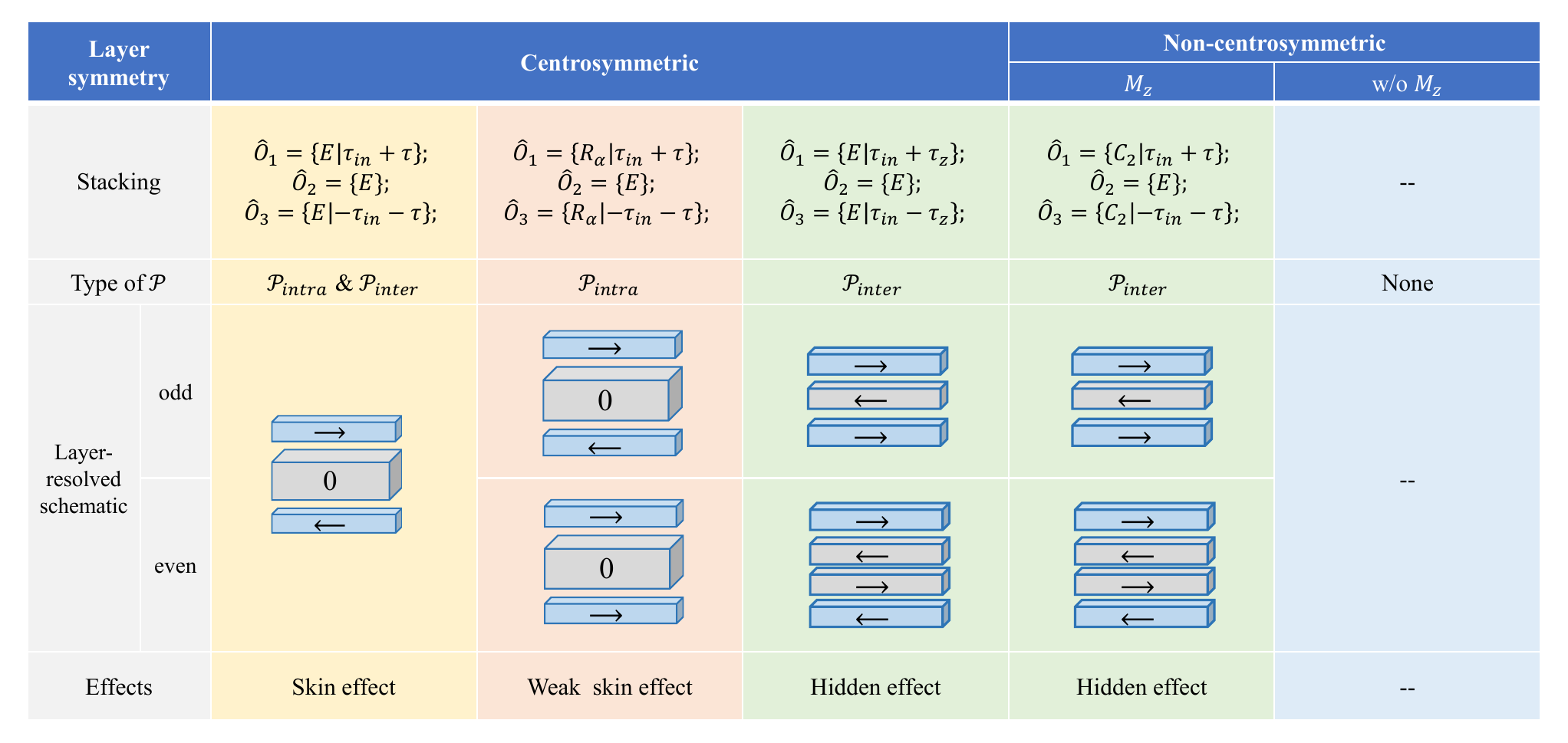} 
	\caption{\label{fig:fig1} The four classes of layer-resolved second-order NLO responses in stacked 2D materials. The classification is based on the position of inversion center, corresponding to 4 possible combinations of $\mathcal{P}_{\text{intra}}$ and $\mathcal{P}_{\text{inter}}$ in stacking process: (i) coexisting $\mathcal{P}_{\text{intra}}$ and $\mathcal{P}_{\text{inter}}$; (ii) pure $\mathcal{P}_{\text{intra}}$; (iii) pure $\mathcal{P}_{\text{inter}}$; (iv) absence of both $\mathcal{P}_{\text{intra}}$ and $\mathcal{P}_{\text{inter}}$. The first three types generate skin effect, weak skin effect, and hidden effect, respectively. If the monolayer is $\mathcal{P}$-broken and lacks mirror symmetry, there is no inversion center in stacked multilayer. The arrows in the schematic indicate the direction of the NLO polarization in each layer.} 
\end{figure*}

\textbf{\emph{Theory of layer-resolved NLO responses.}}
The NLO responses in vdW multilayers are primarily governed by the intrinsic symmetry of individual layers and subsequently modified by interlayer couplings \cite{wangGeSe2017,shanStackingSymmetryGoverned2018,yeNLO2023}. To keep the discussion compact, we begin with the simplest scenario in which each monolayer is nonmagnetic and centrosymmetric. In this case, the SHG and BPVE are rigorously forbidden within each isolated layer in the absence of interlayer coupling. 

Since interlayer coupling in vdW multilayers mainly arises from charge overlap along the out-of-plane direction and is therefore short-ranged, we formulate an effective Hamiltonian that includes only nearest-neighbor interlayer coupling between adjacent layers. This leads to a block-tridiagonal Hamiltonian of the form
\begin{equation}
	H =
	\begin{pmatrix}
		H_{11} & V_{12} & 0      & 0      & 0      & \cdots \\
		V_{21} & H_{22} & V_{23} & 0      & 0      & \cdots \\
		0      & V_{32} & H_{33} & V_{34} & 0      & \cdots \\
		0      & 0      & V_{43} & H_{44} & V_{45} & \cdots \\
		0      & 0      & 0      & V_{54} & H_{55} & \cdots \\
		\vdots & \vdots & \vdots & \vdots & \vdots & \ddots
	\end{pmatrix},
\end{equation}
where $H_{ii}$ represents the Hamiltonian of the $i$th isolated layer and $V_{ij}$ ($i \neq j$) denotes the interlayer coupling between layers $i$ and $j$. In particular, for a surface (outermost) layer, we formally regard it as being sandwiched between a hypothetical vacuum layer on one side and a real material layer on the other, such that the coupling to the vacuum side vanishes identically. To capture the layer-resolved distribution of NLO response, we focus on a minimal local Hamiltonian involving a target layer (e.g., layer 2) and its nearest neighbors, which takes the form
\begin{equation}
	H' =
	\begin{pmatrix}
		H_{11} & V_{12} & 0 \\
		V_{21} & H_{22} & V_{23} \\
		0 & V_{32} & H_{33}
	\end{pmatrix},
\end{equation}
where the target layer is influenced only by its adjacent layers. By formally integrating out the neighboring layers within this trilayer subspace, we obtain an effective monolayer Hamiltonian for layer 2,
\begin{equation}
	H_2^{\mathrm{eff}}(k)
	= H_{22}
	- \left( V_{21} H_{11}^{-1} V_{12}
	+ V_{23} H_{33}^{-1} V_{32} \right).
\end{equation}

Since $H_{22}$ is assumed to be centrosymmetric, the emergence of a nonvanishing NLO response depends on whether the coupling-induced correction term $V_{21} H_{11}^{-1} V_{12} + V_{23} H_{33}^{-1} V_{32}$ breaks inversion ($\mathcal{P}$) symmetry. Notably, as the effective Hamiltonian is constructed locally from three neighboring layers, $\mathcal{P}$ symmetry here refers to that of the corresponding trilayer subsystem. If the stacking order admits a local operator acting on the trilayer that exchanges layers 1 and 3 while leaving layer 2 invariant, then $\mathcal{P}$ symmetry is preserved for layer 2, and its layer-resolved NLO responses must vanish. Consequently, in such stacked multilayers, the NLO response is strongly suppressed in the bulk-like interior. In contrast, for surface layers, $\mathcal{P}$ symmetry is inevitably broken, as a hypothetical vacuum layer can never be mapped onto a real material layer. As a result, a finite NLO response is generically expected in the two outermost layers. This picture establishes the essential physics of the skin effect in 2D vdW multilayers: it arises from stacking-induced symmetry breaking encoded in the effective Hamiltonian, which is a bulk property, rather than originating from surface-specific chemical effects.

In principle, a complete classification of layer-resolved NLO responses can be obtained by progressively breaking the $\mathcal{P}$ symmetry of the effective Hamiltonian in Eq.~(3), as illustrated in Fig.~\ref{fig:fig1}. Before presenting a detailed analysis, however, we emphasize a crucial point: in vdW multilayers there exist two distinct types of inversion symmetry, depending on whether the inversion operation leaves a given layer invariant while exchanging its two neighbors, or instead directly exchanges two adjacent layers\cite{geimRiseGraphene2007,luigrapheneStack2011,liMoS2hBN2013,huangCrI32017}. We denote these two symmetries as $\mathcal{P}_{\mathrm{intra}}$ and $\mathcal{P}_{\mathrm{inter}}$, respectively, corresponding to inversion centers located within an individual layer or between two adjacent layers. We note that $\mathcal{P}_{\mathrm{intra}}$ and $\mathcal{P}_{\mathrm{inter}}$ are neither mutually exclusive nor dependent. They may coexist in a vdW multilayer, exist independently of each other, or be simultaneously absent.

As discussed in Eq.~(3), $\mathcal{P}_{\mathrm{intra}}$ alone is sufficient to give rise to the skin effect, whereby only the top and bottom layers contribute to the NLO response. If $\mathcal{P}_{\mathrm{intra}}$ is broken while $\mathcal{P}_{\mathrm{inter}}$ is preserved, each layer can host a nonvanishing NLO response. However, the $\mathcal{P}_{\mathrm{inter}}$ symmetry enforces equal-magnitude but opposite-sign NLO responses on adjacent layers. According to our theory, the layer-resolved NLO responses in vdW multilayers can be classified according to specific combinations of the presence or absence of $\mathcal{P}_{\mathrm{intra}}$ and $\mathcal{P}_{\mathrm{inter}}$. Correspondingly, the three prototypical forms of layer-resolved NLO responses are illustrated in Fig.~\ref{fig:fig1}.
\begin{enumerate}[label=(\alph*)]
 \item Skin effect ($\mathcal{P}_{\mathrm{intra}}\,\mathbin{\&}\,\mathcal{P}_{\mathrm{inter}}$): 
       The NLO response is significant only at the outermost layers, and the NLO coefficients of the top and bottom layers are symmetry-protected to be antiparallel, i.e., equal in magnitude and opposite in sign.
 
 \item Weak skin effect ($\mathcal{P}_{\mathrm{intra}}$): 
       As in the skin effect, the NLO response is localized at the outermost layers. However, the NLO responses of the top and bottom layers can be either parallel or antiparallel, depending on the parity of the layer number. For an even number of layers, in the absence of $\mathcal{P}_{\mathrm{inter}}$, the NLO polarizations on the top and bottom layers generally differ in magnitude, unless constrained by additional symmetries, such as a mirror symmetry relating the two layers.
       
 \item Hidden effect ($\mathcal{P}_{\mathrm{inter}}$): 
       The NLO polarizations are staggered between adjacent layers, with equal magnitudes and opposite signs, resulting in a vanishing overall NLO response. This behavior is analogous to the hidden spin texture effect \cite{zhangHiddenSpin2014,liuDesignhidden2015}. Owing to the modified interlayer couplings at the surfaces, the NLO magnitudes of the top and bottom layers may differ from those of the inner layers; however, such deviations are purely quantitative and do not alter the staggered pattern enforced by $\mathcal{P}_{\mathrm{inter}}$ symmetry. Stacked non-centrosymmetric layers also belong to this class.
 \end{enumerate}
We note that there exist no chirality-preserving layer operators capable of introducing either $\mathcal{P}_{\mathrm{intra}}$ or $\mathcal{P}_{\mathrm{inter}}$ in a multilayer when the isolated monolayer lacks both inversion and mirror symmetries. Hence, neither the skin effect nor the hidden effect can occur in such systems. The layer-dependent NLO responses are then mainly determined by the intrinsic monolayer NLO response and the stacking sequence.

From a structural perspective, any multilayer can be viewed as being generated from an isolated monolayer by repeatedly applying certain spatial operations that create its neighboring layers. Clarifying this correspondence enables the symmetry classification developed above to be directly connected to geometric operators of vdW multilayers. We now proceed to explicitly identify the layer operations in vdW multilayers that give rise to $\mathcal{P}_{\mathrm{intra}}$ and $\mathcal{P}_{\mathrm{inter}}$.

In space group theory, a standard notation for a layer-generating operator, namely, a spatial operation that relates two neighboring layers, can be written as
\begin{equation}
	\hat{O} = \{ A \,|\, \boldsymbol{\tau} \},
\end{equation}
where $A$ denotes a point-group operation and $\boldsymbol{\tau}$ is the translation vector\cite{dresselhaus2008}. Here, only chirality-preserving operations are considered, including the identity $E$, rotations $R$, and translations $\boldsymbol{\tau}$. Following the trilayer effective Hamiltonian in Eq.~(3), we first consider the case where the monolayer is centrosymmetric and assume three representative layer-generating operators,
$\hat{O}_1 = \{ R_\alpha \mid {\tau}_{\mathrm{in}} + \tau_z \}$,
$\hat{O}_2 = \{ E \mid 0\}$, and
$\hat{O}_3 = \{ R_\beta \mid {\tau}'_{\mathrm{in}} - \tau_z \}$.
By acting these operators on the spatial coordinates of the monolayer $(x, y, z)$ and its centrosymmetric counterpart $(-x, -y, -z)$, all symmetry-equivalent sites of a trilayer slab can be generated. By examining the preservation or breaking of $\mathcal{P}_{\mathrm{intra}}$ and $\mathcal{P}_{\mathrm{inter}}$, any layer-generating operator $\hat{O}$ can then be classified into a specific category shown in Fig.~\ref{fig:fig1}. Detailed classification of layer-stacking operators, as well as those for the non-centrosymmetric systems, can be found in the End Matter.

Furthermore, the present analysis can be generalized to magnetic materials by including the spin degree of freedom. For the magnetic injection current, $\mathcal{P}_{\mathrm{intra}}$ still guarantees the skin effect. However, the spatial profile of the NLO response also depends on the underlying magnetic configuration. The details are provided in the End Matter and the associated Note S1 of Supplemental Material (SM).

\textbf{\emph{Defining layer-resolved NLO coefficients.}}
The BPVE refers to the generation of a direct current under optical excitation in noncentrosymmetric materials, which can be expressed as $\mathbf{J} = 2\boldsymbol{\sigma} : \mathbf{E}\mathbf{E}^*$, where $\boldsymbol{\sigma}$ is the BPVE conductivity tensor. It is well established that the total rectification current can be decomposed into four contributions: the normal shift current (NSC), normal injection current (NIC), magnetic shift current (MSC), and magnetic injection current (MIC)\cite{sipeNLO2000,wangformular2020,chenBasicFormula2022}. In this work, we focus on linearly polarized light, for which only the NSC and MIC are allowed in nonmagnetic and spin-polarized systems, respectively. By applying the layer projection operator $\hat{P}_L = \sum_{l \in L} |\psi_{l}\rangle\langle \psi_{l}|$ onto the Bloch states in the related matrix elements \cite{zhouSkinEffectNonlinear2024}, such as the interband dipole $r_{nm;a}^c$ and the intraband velocity $v_{nn}^a$, the layer-resolved NSC and MIC coefficients are obtained as
\begin{multline}
	\sigma_{L}^{abc} = -\frac{i \pi e^3}{4 \hbar^2} \int [d\mathbf{k}]\sum_{n,m,\sigma} f_{nm} (r_{mn}^b r_{nm;a,L}^c + r_{mn}^c r_{nm;a,L}^b) \\
	\times [\delta(\omega_{nm} - \omega)+\delta(\omega_{mn} - \omega)],
\end{multline}
\begin{multline}
	\eta_{L}^{abc} = -\frac{\pi e^3}{2 \hbar^2} \int [d\mathbf{k}] \sum_{n,m,\sigma} f_{mn}\Delta_{mn,L}^a \{r_{nm}^b,r_{mn}^c\} \\
	\times \delta(\omega_{nm} -\omega),
\end{multline}
where $|\psi_{l}\rangle$ is the local atomic orbital basis in the Wannier representation and $L$ denotes layer number. The detailed derivation of layer-projected formula is provided in the Note S2 of SM.

\begin{figure*}[!t]
	\includegraphics[width=0.9\textwidth]{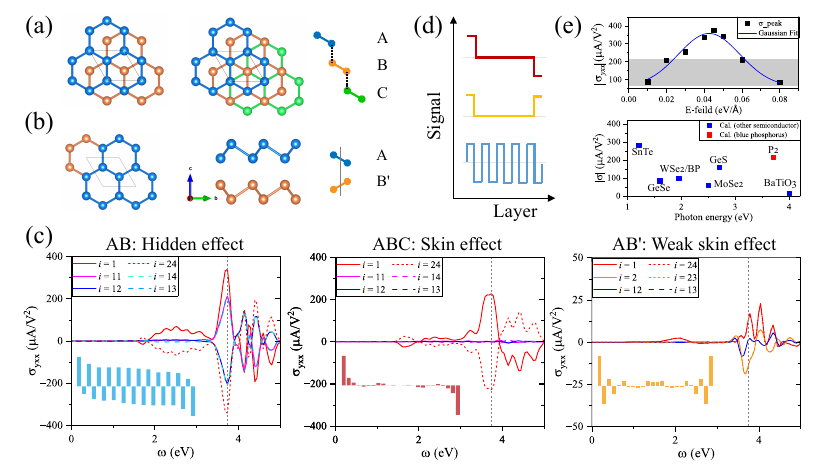} 
	\caption{\label{fig:fig2} (a) The schematic of AB and ABC stacking sequences, where each layer laterally shifted by $(1/3,-1/3,0)$. (b) The schematic of AB$^\prime$ stacking sequence via $C_2$ screw symmetry. (c) Layer-resolved shift current coefficient $\sigma^{yxx}$ of a 24-layers blue phosphorus slab with AB, ABC and AB$^\prime$ stacking. (d) Concept of a multi-channel signal abstracted from layer-resolved NLO responses. The values of responses are mapped to 1, 0, or -1, based on whether they are positive, zero, or negative. (e) Upper panel: electric-field-modulated $\sigma^{yxx}$ of monolayer blue phosphorus. The gray region denotes the range of $\sigma^{yxx}$ values arising from different stacking orders in (c). Lower panel: comparison between the NSC coefficients on a surface layer of ABC-stacked blue phosphorus and the intrinsic responses of other monolayer materials reported in the literature \cite{young2012first,rangel2017large,wang2019ferroicity,akamatsu2021van,jin2024peculiar}.} 
\end{figure*}

\textbf{\emph{Calculations on blue phosphorus and CrI$_3$.}}
To elucidate the stacking-order-dependent layer-resolved NLO responses in nonmagnetic systems, we take 2D blue phosphorus as an example. Its monolayer is centrosymmetric, belonging to the space group $P\bar{3}m1$. According to the symmetry criteria established above, we construct three distinct stacking sequences, namely AB, ABC, and AB$^\prime$, as illustrated in Figs.~\ref{fig:fig2}(a) and \ref{fig:fig2}(b).

The first two stacking sequences, AB and ABC, are generated by pure translational operations, in which each layer is laterally shifted by $(1/3,-1/3,0)$ relative to the previous one, but with different stacking periodicities of two and three atomic layers, respectively. In the AB stacking, the inversion center exists only in the vdW gap, corresponding to the presence of $\mathcal{P}_{\mathrm{inter}}$ symmetry, which leads to the staggered NLO hidden effect.  
By contrast, the ABC stacking preserves both $\mathcal{P}_{\mathrm{intra}}$ and $\mathcal{P}_{\mathrm{inter}}$ symmetries, supporting the skin effect characterized by antiparallel NLO polarizations on the top and bottom layers. 
Regarding the AB$^\prime$ stacking, adjacent layers are rotated by $180^\circ$ with respect to each other about an axis normal to the layers and passing through the inversion center. Thus, the inversion center exists only within individual layers, corresponding to the presence of $\mathcal{P}_{\mathrm{intra}}$ symmetry, which gives rise to the weak skin effect. Indeed, AB$^\prime$-stacked multilayers with an even number of layers are globally $\mathcal{P}$-broken. In this case, the parallel NLO responses of the top and bottom layers are related by the out-of-plane mirror symmetry. The layer-dependent space groups and the corresponding NLO response types for the three stacking configurations are summarized in the Table S1 of SM.

Figure~\ref{fig:fig2}(c) shows the layer-resolved NSC coefficients $\sigma^{yxx}_{L}$ of a 24-layer blue phosphorus multilayer in AB, ABC, and AB$^\prime$ stacking configurations. Convergence with respect to the number of layers is shown in Fig. S2 of SM. As expected, the AB stacking exhibits a hidden effect characterized by staggered layer-to-layer NSC coefficients, particularly deep inside the slab. For instance, at the pronounced peak around $\hbar\omega = 3.7$~eV, $\sigma^{yxx}_{L}$ takes the values 212.7, $-204.6$, 204.1, and $-213.2$ for layers 11 to 14, respectively. When the stacking sequence changes from AB to ABC, the NSC response in the bulk-like region is completely suppressed due to the presence of local $\mathcal{P}_{\mathrm{intra}}$ symmetry, giving rise to a skin effect. In contrast to the ABC stacking, where the $\sigma^{yxx}_{L}$ coefficients on the top and bottom layers are antiparallel, the NSC responses of the two outermost layers are identical in the AB$^\prime$ stacking. This behavior corresponds to the weak skin effect and is further protected by the out-of-plane $M_z$.  
If we draw an analogy between layer-resolved responses and multi-channel signal, the toggling of layer signals among the states 1, 0, and -1 becomes readily conceptualized, as shown in Fig.~\ref{fig:fig2}(d). We note that, in Fig.~\ref{fig:fig2}(e), the stacking-induced response reaches a magnitude comparable to that achieved through electric-field tuning and is also comparable to the intrinsic responses reported in other 2D materials. This highlights stacking as a practical route to sizable NLO responses without external fields, making it promising for surface-selective and integrated devices. Further, the overall strength of layer-resolved NLO responses can be tuned by the interlayer spacing (Figs. S3 and S4 in SM). 

\begin{figure}[htbp]
	\centering
	\includegraphics[width=\columnwidth]{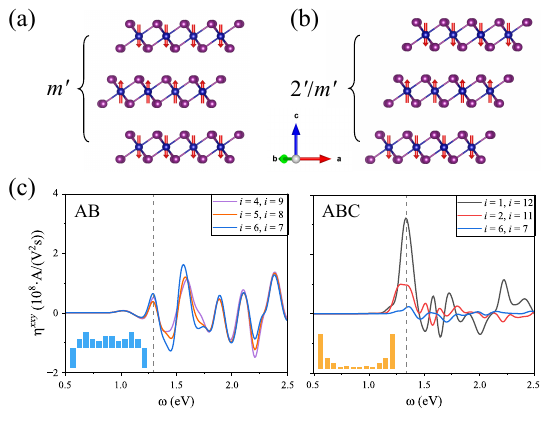} 
	\caption{\label{fig:fig3} Lattice structures of triple-layer A-type antiferromagnetic CrI$_3$ in (a) AB and (b) ABC stacking. The blue and purple balls represent the Cr and I atoms. The red arrow denotes the orientation of the local spin moments. Note that the structure should be doubled to recover the full periodicity due to the interlayer antiferromagnetic order. (c) The calculated layer-resolved MIC coefficient $\eta^{xxy}$ of 12-layer CrI$_3$ slabs in AB (left) and ABC (right) stacking.} 
\end{figure}

Next, we investigate the stacking-order dependence of the layer-resolved MIC responses in A-type antiferromagnetic CrI$_3$ \cite{jiangStackingCrI32019,chenObservCrI32019,songSwitching2DCrI32019}. Figures~\ref{fig:fig3}(a) and \ref{fig:fig3}(b) show the local atomic structures of AB- and ABC-stacked 12-layer CrI$_3$ slabs with A-type spin alignment. For both stacking configurations, the magnetic point group of the entire slab is $\mathrm{2/}m^{\prime}$, which allows for nonvanishing MIC tensor elements, including $xxy$, $yxx$, and $yyy$. Despite sharing the same global magnetic point group, the local symmetry experienced by individual layers differs, leading to distinct layer-resolved MIC responses (see End Matter). Specifically, the magnetic point groups of the ABA and ABC trilayer building blocks can be decomposed as $\{E\}\oplus \mathcal{T}\{M_y\}$ and $\{E,I\}\oplus \mathcal{T}\{M_y,C_{2y}\}$, respectively. In the AB stacking, the breaking of $\mathcal{P}_{\mathrm{intra}}$ suppresses the skin effect, while the preserved $\mathcal{P}_{\mathrm{inter}}\mathcal{T}$ enforces identical MIC coefficients across all layers (except for the surfaces), resulting in the uniform effect. By contrast, in the ABC stacking, the inner layers simultaneously preserve local $\mathcal{P}_{\mathrm{intra}}$ and $\mathcal{P}_{\mathrm{inter}}\mathcal{T}$ symmetries, giving rise to a pronounced skin effect. As shown in Fig.~\ref{fig:fig3}(c), the MIC responses of layers 4-9 exhibit nearly identical spectral line shapes over the entire frequency range in the AB-stacked configuration. By contrast, in the ABC-stacked CrI$_3$, the MIC coefficient $\eta^{xxy}$ is strongly suppressed from the outermost layers toward the interior, consistent with the skin effect picture. Thus, we have provided a detailed analysis of the layer-resolved MIC in A-type antiferromagnets. In principle, the symmetry-based framework developed here can be extended to more complex magnetic configurations, such as G-type antiferromagnets.

\textbf{\emph{Conclusion.}} 
We have developed a general theoretical framework for classifying layer-resolved second-order NLO responses in vdW materials. By analyzing local inversion symmetries, we identify three distinct types of nonuniform layer distributions: the skin effect, the weak skin effect, and the hidden effect, which arise from specific combinations of intralayer and interlayer inversion centers. We further establish a direct correspondence between stacking operations and the symmetry of resulting layer sequences. Moreover, the framework naturally extends to magnetic systems, providing new insights into magnetic BPVE and related magnetic NLO coefficients. More broadly, as a general theoretical language, this paradigm is applicable to other physical effects that are sensitive to centrosymmetry, such like the nonlinear Hall effect \cite{maObservNLHall2019,sodemannNLHall2015}. Our results offer a unified understanding of surface-localized and spatially nonuniform NLO response in layered materials, and open new avenues for engineering optoelectronic functionalities through stacking geometry and symmetry control.

\section{Acknowledgments}
This work was supported by the National Natural Science Foundation of China (NSFC) (Grant Nos. W2511008 and 12088101) and the Science Challenge Project (Grant No. TZ2025013).

\bibliographystyle{apsrev4-2-title}
\bibliography{refs}

\section{End Matter}
\textbf{\emph{General classification of layer-stacking operator.}} 
To investigate the local symmetry of a target layer in multilayer slabs, we construct a trilayer slab model including only nearest-neighbor interlayer coupling. The trilayer slab consists of the target layer and its two adjacent layers. In particular, for a surface layer, we assume it is sandwiched between a hypothetical vacuum layer and the immediately adjacent inner layer. The stacking sequence determines the local symmetry of the target layer. 

To describe the stacking sequence, we adopt a notation for layer operations from space-group theory: $\hat{O}=\{R_{\alpha}|\tau\}$, where $R_{\alpha}$ and $\tau$ denote a point-group operation and a translation vector, respectively\cite{dresselhaus2008}.We then introduce three layer operations: $
\hat{O}_1=\{R_\alpha\mid \tau_{\rm in}+\tau_z\},\quad
\hat{O}_2=\{E\mid 0\},\quad
\hat{O}_3=\{R_\beta\mid \tau_{\rm in}'-\tau_z\}.$ 
Applying these operations to the inversion-related coordinates of a centrosymmetric monolayer, $(x,y,z)$ and $(-x,-y,-z)$, generates all symmetry-related sites in the trilayer slab. The action of a Seitz operation on the coordinates can be written as
\begin{equation}
    \begin{pmatrix}
        1 & 0 \\
        \tau & R_{\alpha}
    \end{pmatrix}
    \begin{pmatrix}
        1 \\
        r
    \end{pmatrix} = 
	\begin{pmatrix}
		1 \\
		\tau + R_{\alpha} \cdot r 
	\end{pmatrix} =
	\begin{pmatrix}
		1 \\
		r'
	\end{pmatrix}
\end{equation}
where the matrix
\begin{equation} 
\begin{pmatrix}
		1 & 0 \\
		\tau & R_{\alpha}
\end{pmatrix}
\end{equation}
is a matrix representation of the space-group operation $\{R_{\alpha}|\tau\}$. Here, $1$ is a number, $0$ denotes a row vector of three zeros, $\tau$ is a column vector representing the translation, and $R_{\alpha}$ is a $3\times 3$ rotation matrix. 

For a centrosymmetric monolayer with inversion-related coordinates $r=(x,y,z)$ and $-r=(-x,-y,-z)$, applying the three layer operations defined above generates all symmetry-related sites in the trilayer slab:
\begin{widetext}
\begin{equation}
\begin{aligned}
\hat{O}_1
\begin{pmatrix}1\\ r\end{pmatrix}
&=
\begin{pmatrix}
1 & 0\\
\tau_{\rm in}+\tau_z & R_{\alpha}
\end{pmatrix}
\begin{pmatrix}1\\ r\end{pmatrix}
=
\begin{pmatrix}
1\\
R_{\alpha}r+\tau_{\rm in}+\tau_z
\end{pmatrix},
&
\hat{O}_1
\begin{pmatrix}1\\ -r\end{pmatrix}
&=
\begin{pmatrix}
1\\
-R_{\alpha}r+\tau_{\rm in}+\tau_z
\end{pmatrix};
\\[0.8em]
\hat{O}_2
\begin{pmatrix}1\\ r\end{pmatrix}
&=
\begin{pmatrix}
1 & 0\\
0 & E
\end{pmatrix}
\begin{pmatrix}1\\ r\end{pmatrix}
=
\begin{pmatrix}
1\\
r
\end{pmatrix},
&
\hat{O}_2
\begin{pmatrix}1\\ -r\end{pmatrix}
&=
\begin{pmatrix}
1\\
-r
\end{pmatrix};
\\[0.8em]
\hat{O}_3
\begin{pmatrix}1\\ r\end{pmatrix}
&=
\begin{pmatrix}
1 & 0\\
\tau_{\rm in}'-\tau_z & R_{\beta}
\end{pmatrix}
\begin{pmatrix}1\\ r\end{pmatrix}
=
\begin{pmatrix}
1\\
R_{\beta}r+\tau_{\rm in}'-\tau_z
\end{pmatrix},
&
\hat{O}_3
\begin{pmatrix}1\\ -r\end{pmatrix}
&=
\begin{pmatrix}
1\\
-R_{\beta}r+\tau_{\rm in}'-\tau_z
\end{pmatrix}.
\end{aligned}
\end{equation}
\end{widetext}
Thus, the six symmetry-related sites in the trilayer slab are
$r$, $-r$, $R_{\alpha}r+\tau_{\rm in}+\tau_z$, $-R_{\alpha}r+\tau_{\rm in}+\tau_z$,
$R_{\beta}r+\tau_{\rm in}'-\tau_z$, and $-R_{\beta}r+\tau_{\rm in}'-\tau_z$. We then consider three representative stacking-induced combinations of $\mathcal{P}_{\rm intra}$ and $\mathcal{P}_{\rm inter}$:
\begin{enumerate}[label=(\alph*)]
\item The system possesses both intralayer and interlayer inversion symmetries. This requires $R_{\alpha}=R_{\beta}=E$ and $\tau_{\rm in}=-\tau_{\rm in}'$.
\item The trilayer slab possesses inversion symmetry, which requires either $R_{\alpha} = R_{\beta}$ and $\tau_{\mathrm{in}} = -\tau_{\mathrm{in}}'$, or $R_{\alpha} = -R_{\beta}$ and $\tau_{\mathrm{in}} = -\tau_{\mathrm{in}}'$.
\item The inversion center lies in the vdW gap, corresponding to $R_{\alpha}=R_{\beta}=E$. For a regular stacking sequence along the $z$ direction, we further take $\tau_{\rm in}=\tau_{\rm in}'\neq 0$.
\end{enumerate}

\begin{figure}
	\centering
	\includegraphics[width=\columnwidth]{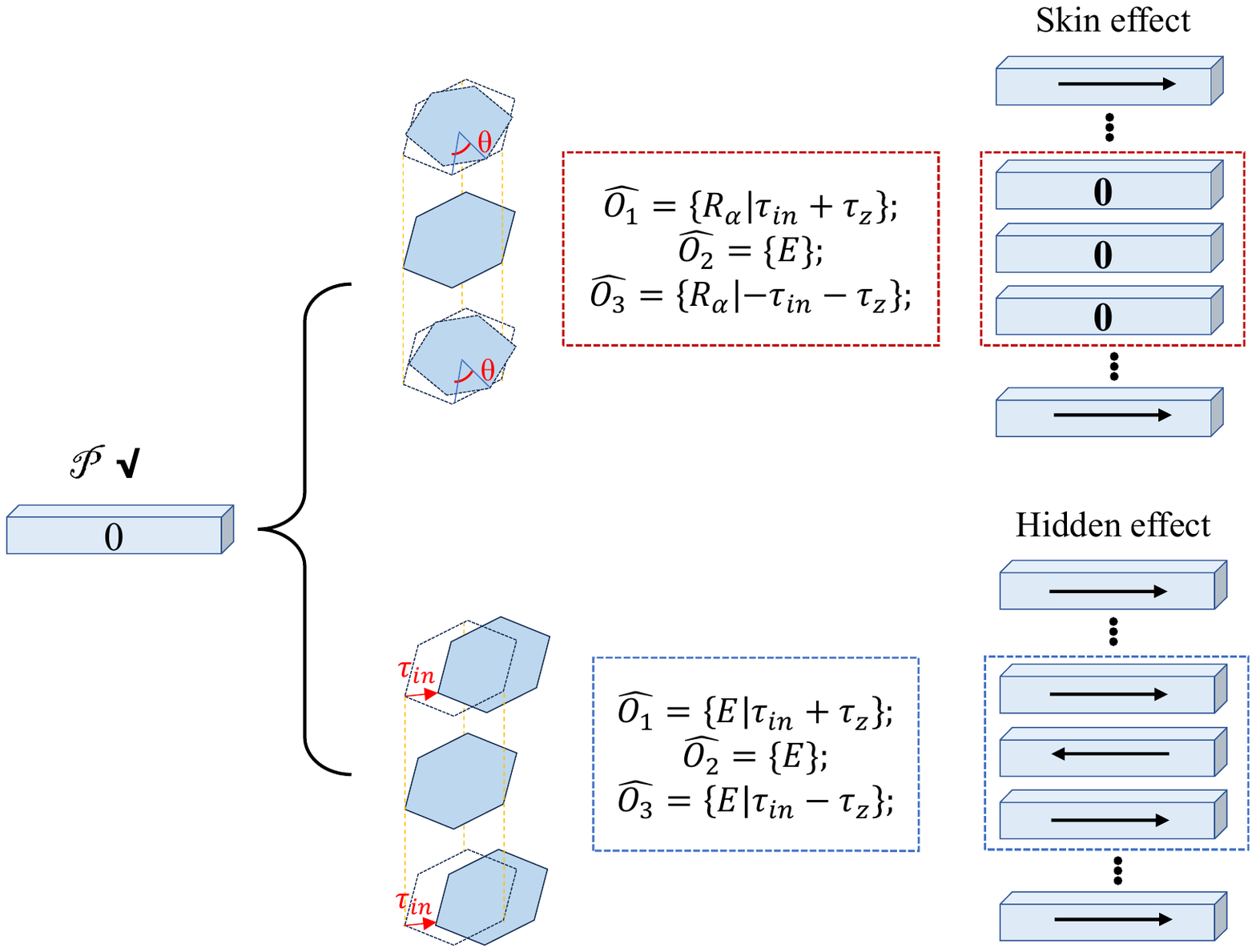}
	\caption{Schematic diagrams of two typical stacking modes and their corresponding layer-resolved NLO responses. }
	\label{fig:S1}
\end{figure}

This framework establishes a direct connection between layer-resolved NLO responses and stacking configurations. For example, preserving both $\mathcal{P}_{\rm intra}$ and $\mathcal{P}_{\rm inter}$ while keeping layer~2 fixed requires the layer-generating operations for layers~1 and~3 to be pure translations. Thus,
$\hat{O}_1=\{E\mid {\tau}_{\rm in}+\tau_z\}$ and
$\hat{O}_3=\{E\mid -{\tau}_{\rm in}-\tau_z\}$,
giving rise to the skin effect.
To break $\mathcal{P}_{\rm inter}$, layers~1 and~3 can be rotated by the same finite angle,
$\hat{O}_1=\{R_\alpha\mid {\tau}_{\rm in}+\tau_z\}$ and
$\hat{O}_3=\{R_\alpha\mid -{\tau}_{\rm in}-\tau_z\}$.
This construction preserves the mapping between layers~1 and~3 and thus keeps $\mathcal{P}_{\rm intra}$ intact. However, it breaks the inversion relation between adjacent layers, leading to the weak skin effect.
Similarly, to break $\mathcal{P}_{\rm intra}$, layers~1 and~3 can be translated along the same in-plane direction, i.e.,
$\hat{O}_1=\{E\mid {\tau}_{\rm in}+\tau_z\}$ and
$\hat{O}_3=\{E\mid {\tau}_{\rm in}-\tau_z\}$.
In this case, the effective mapping center relating layers~1 and~3 no longer coincides with the inversion center of layer~2, thereby breaking $\mathcal{P}_{\rm intra}$ and giving rise to the hidden effect. Figure~\ref{fig:S1} schematically illustrates two representative stacking configurations and their corresponding layer-resolved NLO responses.

Finally, for a noncentrosymmetric monolayer with out-of-plane mirror symmetry $M_z$, $\mathcal{P}_{\rm intra}$ is necessarily broken, whereas $\mathcal{P}_{\rm inter}$ can, in principle, be preserved. This generally requires a specific in-plane rotation, such as a twofold rotation $C_2$, giving
$\hat{O}_1=\{C_2\mid {\tau}_{\rm in}+\tau_z\}$ and
$\hat{O}_3=\{C_2\mid -{\tau}_{\rm in}-\tau_z\}$.
This configuration therefore gives rise to the hidden effect.

\begin{figure*}[!t]
	\centering
	\includegraphics[width=\textwidth]{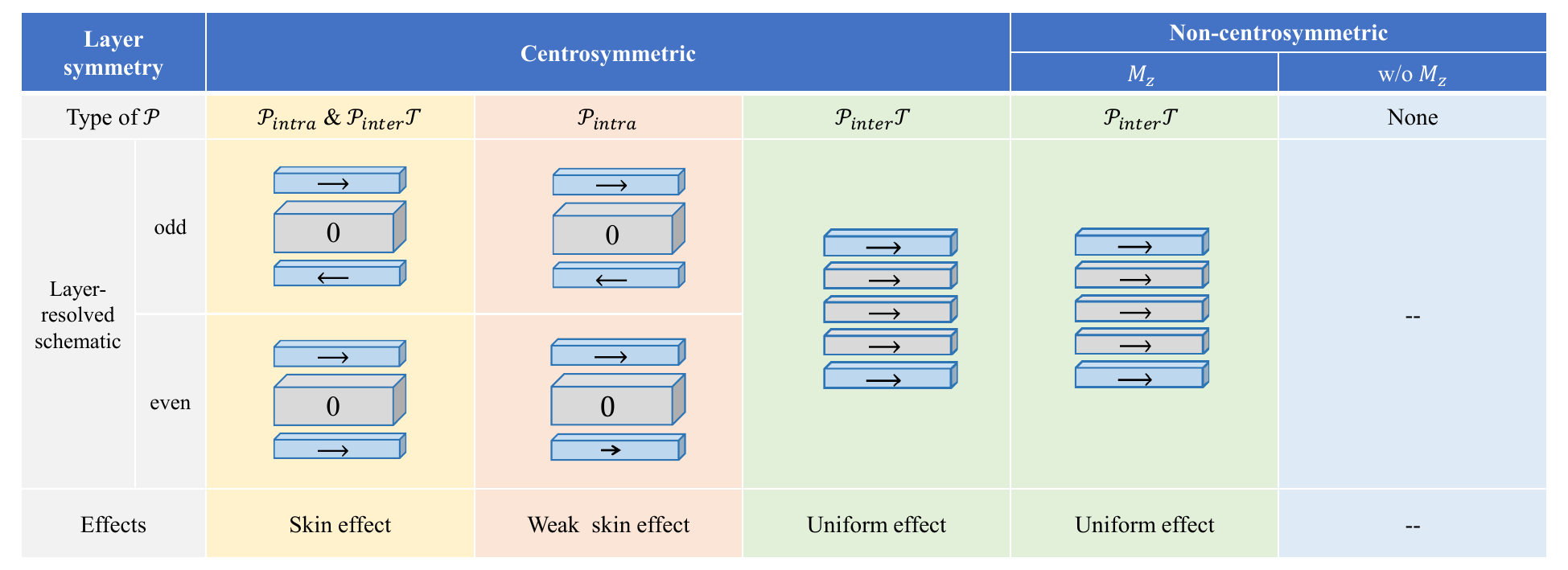} 
	\caption{\label{fig:magnetica-table} Distribution of layer-resolved MIC coefficients in A-type antiferromagnets based on stacking. Here, $\mathcal{P}_{\text{intra}}$ represents that the inversion center is located at the intralayer, which preserves the skin effect; $\mathcal{P}_{\mathrm{inter}}\mathcal{T}$ denotes a symmetry operation where the spatial inversion center lies within the interlayer space and must be combined with time reversal.}
\end{figure*}

\textbf{\emph{Layer-resolved NLO responses in magnetic materials.}} 
Here we take the A-type interlayer antiferromagnet as a representative example to illustrate the possible layer-resolved MIC responses, as shown in Fig.~\ref{fig:magnetica-table}. In this magnetic configuration, the spin moments are aligned parallel within each layer but antiparallel between adjacent layers. It can be seen that the skin effect guaranteed by $\mathcal{P}_{\mathrm{intra}}$, persists in the antiferromagnetic case, regardless of whether $\mathcal{P}_{\mathrm{inter}}\mathcal{T}$ is present. On the other hand, the presence or absence of $\mathcal{P}_{\mathrm{inter}}\mathcal{T}$ determines whether the skin effect responses of the top and bottom layers are symmetry-protected. When only $\mathcal{P}_{\mathrm{inter}}\mathcal{T}$ is preserved but $\mathcal{P}_{\mathrm{intra}}$ is broken, the system exhibits a constant layer-resolved MIC distribution, which we refer to as the uniform effect. This spatial distribution pattern stands in sharp contrast to the staggered response found in nonmagnetic systems with only $\mathcal{P}_{\mathrm{inter}}$. More details of the derivation are given in the Note S1 of SM.

\end{document}